# Possible helimagnetic order in Co$^{4+}$-containing perovskites Sr$_{1-x}$Ca$_x$CoO$_3$

Hidefumi Takahashi,[1,2,3] Masaho Onose,[2] Yasuhito Kobayashi,[1] Takahiro Osaka,[2] Soushi Maeda,[1] Atsushi Miyake,[4] Masashi Tokunaga,[4] Hajime Sagayama,[5] Yuichi Yamasaki,[6] and Shintaro Ishiwata[1,2,3]

[1]*Division of Materials Physics and Center for Spintronics Research Network (CSRN), Graduate School of Engineering Science, Osaka University, Toyonaka, Osaka 560–8531, Japan*
[2]*Department of Applied Physics, The University of Tokyo, Tokyo 113–8656, Japan*
[3]*Spintronics Research Network Division, Institute for Open and Transdisciplinary Research Initiatives, Osaka University, Yamadaoka 2-1, Suita, Osaka, 565-0871, Japan*
[4]*The Institute for Solid State Physics, The University of Tokyo, Kashiwa, Chiba 277-8581, Japan*
[5]*Institute of Materials Structure Science (IMSS), High Energy Accelerator Research Organization (KEK), Tsukuba, Ibaraki 305-0801, Japan*
[6]*National Institute for Materials Science (NIMS), Tsukuba, Ibaraki 305-0047, Japan*

We systematically synthesized perovskite-type oxides Sr$_{1-x}$Ca$_x$CoO$_3$ containing unusually high valence Co$^{4+}$ ions by a high pressure technique, and investigated the effect of systematic lattice change on the magnetic and electronic properties. As the Ca content $x$ exceeds about 0.6, the structure changes from cubic to orthorhombic, which is supported by the first-principles calculations of enthalpy. Upon the orthorhombic distortion, the ground state remains to be apparently ferromagnetic with a slight drop of the Curie temperature. Importantly, the compounds with $x$ larger than 0.8 show antiferromagnetic behavior with positive Weiss temperatures and nonlinear magnetization curves at lowest temperature, implying that the ground state is noncollinear antiferromagnetic or helimagnetic. Considering the incoherent metallic behavior and the suppression of the electronic specific heat at high $x$ region, the possible emergence of a helimagnetic state in Sr$_{1-x}$Ca$_x$CoO$_3$ is discussed in terms of the band-width narrowing and the double-exchange mechanism with the negative charge transfer energy as well as the spin frustration owing to the next-nearest neighbor interaction.

Searching for novel magnetic phases in a simple system is an attractive yet challenging task in condensed matter physics. One promising approach to discovering exotic magnetic phases is to focus on the system with competing exchange interactions. Whereas this is typified by antiferromagnetic insulators with geometrically frustrated spin lattice[1], frustrated magnetic metals with competing exchange interactions have attracted attention for their novel spintronic phenomena, as highlighted by the emergence of magnetic skyrmion lattice in noncentrosymmetric magnets with Dzyaloshinskii-Moriya interaction[2,3].

As centrosymmetric counterparts of frustrated magnetic metals, perovskite-type oxides with unusually high valence $Fe^{4+}$, $Co^{4+}$, and $Ni^{3+}$ ions have been known to show rich magnetic phases[4–6]. This is exemplified by the observation of various topological spin orders involving multiple-Q helimagnetic (HM) states in $SrFeO_3$[7–10] and a room-temperature ferromagnetic (FM) order in $SrCoO_3$.[11–16] Although they exhibit apparently different magnetic behavior, there are important similarities between them; both of them share the cubic perovskite-type structure, the strong $p$-$d$ hybridization with a negative charge-transfer energy, and the $e_g^1$ configuration, provided that $Co^{4+}$ is in the unique intermediate spin IS state ($t_{2g}^4 e_g^1$) which has been under debate.[11–15,17,18] Considering the fact that the negative charge transfer energy and the $e_g^1$ configuration have been discussed as important conditions for the emergence of the spin spiral in $SrFeO_3$ [6], it is tempting to expect the emergence of the novel HM phase in the $Co^{4+}$-containing perovskites as well.

So far, the $Fe^{4+}$-containing perovskites $AFeO_3$ have been systematically synthesized to study the A-site dependent variation of the magnetic ground state[5,19–22]. Substitution of the A-site in $SrFeO_3$ results in a significant change in the helimagnetic propagation vector in $BaFeO_3$ and $CaFeO_3$, the latter of which shows a metal-insulator transition accompanied by charge disproportionation of $Fe^{4+}$.[5,19–21] On the other hand for the $Co^{4+}$-containing perovskites $ACoO_3$, $Sr_{1-x}Ba_xCoO_3$ and $CaCoO_3$ have been successfully synthesized by a high pressure technique[23–25]. Whereas $Sr_{1-x}Ba_xCoO_3$ has been reported to retain the cubic structure and show the change in the ground state from FM to HM at around $x = 0.4$, there remains a missing link between cubic $SrCoO_3$ with the FM state and orthorhombically distorted $CaCoO_3$ showing antiferromagnetic behavior.

In this paper, we have systematically synthesized perovskite-type cobaltates $Sr_{1-x}Ca_xCoO_3$ by a high-pressure technique and investigated change in the structure and the magnetic states while revealing the structure-property relationship. As the Ca content $x$ increases, the structure changes from cubic to orthorhombic at around $x = 0.6$ and the magnetic ground state changes from FM to AFM at around $x = 0.8$ in the orthorhombic phase. We discuss the origin of variation in the magnetic ground states and the possibility of the helical spin order in the AFM state in terms of the band-width narrowing and the competing exchange interactions inherent to the orthorhombic distortion.

Polycrystalline samples of $Sr_{1-x}Ca_xCoO_3$ were prepared by following the procedure reported in Ref. 23. The starting materials, $SrCO_3$, $CaCO_3$ and $Co_3O_4$, were stoichiometrically mixed. The mixture was heated at 1173 K for 12 h in air. The obtained powder was pelletized and sintered again at 1373 K for 24 h in a flow of oxygen (1 atm), followed by quenching into water at room temperature. Through these processes, oxygen-deficient perovskites $Sr_{1-x}Ca_xCoO_{3-\delta}$ were synthesized. To obtain fully oxidized compounds, the quenched pellet was pulverized and packed into a gold capsule together with an oxidizer $NaClO_3$. The capsule was annealed at 753 K for 1 h at a high pressure of 8 GPa using a cubic-anvil-type high-pressure apparatus. Note that the oxidation of the compound with $x=0.2$ was completed by the dip in an aqueous solution of sodium hypochlorite for 12 h, which turned out to yield a more oxidized sample as compared with the samples prepared by the high-pressure process. The synchrotron powder x-ray diffraction (XRD) with a wavelength of 0.68975 Å was carried out at BL-8A, Photon Factory, KEK, Japan. The diffraction patterns were analyzed by the Rietveld refinement using RIETAN-FP.[26] The magnetization $M$, electrical resistivity $\rho$, and specific heat $C$ were measured using MPMS and PPMS manufactured by Quantum Design, respectively. The magnetization up to 55 T was measured using the nondestructive pulsed magnet with a pulse duration of 36 ms at the International MegaGauss Science Laboratory at the Institute for Solid State Physics.

To evaluate the structural stability of $SrCoO_3$ and $CaCoO_3$ under ambient and high pressures, we performed DFT calculations using the plane-wave-basis projector augmented wave (PAW) method[27] with GGA-PBEsol approximations, as implemented in Quantum Espresso package[28]. We performed structural optimization calculations on cubic and orthorhombic perovskite-type structures at 0 and 10 GPa. Nonmagnetic (NM) and ferromagnetic (FM) configuration were assumed for Co ions, and the electron correlation $U$ was set to 5 eV. The cutoff energies for wavefunctions and charge density were set to 60 Ry and 445 Ry, respectively. The Monkhorst–Pack $k$-point meshes of $9 \times 9 \times 9$ and $6 \times 4 \times 6$ or more were adopted for the cubic and the orthorhombic structures, respectively.

Figure 1(a) shows magnified powder XRD patterns of $Sr_{1-x}Ca_xCoO_3$ with selected compositions of $x$=0.2, 0.4, 0.6, 0.7, 0.8, 1.0. The diffraction patterns for $x = 0.2, 0.4$ and $x= 0.8, 1.0$ were indexed with a cubic unit cell (S.G.: *Pm-3m*) and a $GdFeO_3$-type orthorhombic unit cell (S.G.: *Pbnm*), respectively. The XRD patterns with Rietveld refinement for the cubic phase with $x = 0.4$ and the orthorhombic phase with $x = 0.8$ can be found in Fig. S1 in Supplementary Materials. The lattice parameters are summarized in Table I, and the refined structures for $x = 0$ and 1.0 are shown in Fig. 1(b). It is noted that the diffraction peaks of $x$=0.6 and 0.7 tend to be diffused, implying that these samples are located around the first-order structural phase boundary and contain both cubic and orthorhombic phases coexisting in a microscopic scale. In Figs. 1(c) and (d), the unit cell volume divided by the number of atoms per unit cell $Z$ and the Co-O-Co bond angle are presented as a function of $x$. The unit cell volume of both structures decreases with increasing $x$, reflecting the smaller ionic radius of $Ca^{2+}$ than that of $Sr^{2+}$. Using the Co-O bond lengths, we estimated the bond-valence sums for Co ions[29], which are in the range of +3.86 ~ 4.08 (see Table I). Thermogravimetric (TG) measurements support this high Co valence; the TG measurements suggest that this system is sufficiently oxidized with little oxygen deficiency, implying that the valence of Co is close to +4 (see Supplementary Information).

To examine the stability of cubic and orthorhombic perovskite structures for $SrCoO_3$ and $CaCoO_3$, respectively, the enthalpies at 0 and 10 GPa were calculated based on *ab*-initio calculations. Figure 2 shows the enthalpy of the orthorhombic phase relative to that of the cubic phase for $SrCoO_3$ and $CaCoO_3$ with ferromagnetic (FM) and nonmagnetic (NM) states. It is notable that the orthorhombic structure tends to be more stable than the cubic structure not only for $CaCoO_3$ but for $SrCoO_3$. Assuming the FM state at a high pressure of 10 GPa, while the orthorhombic structure remains to be stable for $CaCoO_3$, the cubic structure becomes more stable than the orthorhombic one for $SrCoO_3$, implying the importance of the FM interaction for the phase formation. This result is qualitatively consistent with our experimental results that the cubic structure found for $SrCoO_3$ is replaced by the orthorhombic structure as the Ca content $x$ increases.

Figure 3(a) shows the temperature dependence of magnetization $M$ divided by a magnetic field $H$ of 0.1 T, $M/H$, measured in field cooling (FC) runs. As $x$ increases from 0 to 0.7, the FM transition temperature $T_c$ decreases systematically, which is accompanied by the emergence of a stepwise transition at $x = 0.6$ and 0.67. The observation of the successive FM transitions reflects the coexistence of cubic and orthorhombic phases with different $T_c$, being consistent with the XRD experiments. For $x = 0.8$ with orthorhombic structure, the FM transition was observed at 120 K, followed by the decrease of $M/H$ below 90 K. Eventually at $x = 1.0$, the FM transition disappeared and instead an antiferromagnetic (AFM) transition was found at 95 K ($=T_N$) as shown in Fig. 3 (b).

Further insight into the magnetic ground state of $Sr_{1-x}Ca_xCoO_3$ is provided by the field dependence of $M$ at 2 K as shown in Fig. 3(c). For $x = 0.2$ and 0.4, the magnetization shows a characteristic field dependence as ferromagnetic $SrCoO_3$. The saturated moment of $\sim 2.0$ $\mu_B$/Co is slightly smaller than that of the single crystal of $SrCoO_3$ ($\sim 2.5$ $\mu_B$/Co)[11], but comparable to that of the polycrystalline sample ($\sim 2.1$ $\mu_B$/Co)[30-32]. The FM behavior was also observed for $x = 0.6$-$0.7$, while the hysteresis loop was enlarged and the saturation moment was slightly diminished probably due to the emergence of the orthorhombic phase. At $x = 0.8$, the $M$-$H$ curve shows a nonlinear and hysteretic behavior, suggesting that an AFM state tends to predominate over the competing FM state as the magnetic ground state. The magnetization curve of $x = 1$ shows AFM-like linear behavior without saturation up to 7 T, unlike the other compounds showing hysteretic and FM-like behavior. To gain more insight into the magnetism of the compound with $x = 1$, we performed magnetization measurements under high fields up to 55 T at various temperatures, as shown in Fig. 3(d). The $M$-$H$ curve at 4.2 K was almost saturated above 40 T with the moment of 1.6 $\mu_B$/Co. The abrupt increase above 9 T can be ascribed to a spin flop transition possibly inherent to the noncollinear or helical spin structure. The magnetization of of $x = 1$ at 40 T is slightly smaller than that of $x = 0.4$, which may be a sign of the partial emergence of the low-spin state ($\sim 1.0$ $\mu_B$/Co). On the other hand, the effective Bohr magnetons number $P_{eff}$ = 3.35-3.98 $\mu_B$ for all compounds estimated from the Curie-Weiss law $M/H(T) = C_w/(T-\theta)$, where $C_w$ is the Curie constant (see Figs. S3 and S4 in Supplementary Materials), is close to the theoretical value of 3.87 $\mu_B$ expected for the intermediate spin state of $Co^{4+}$ ($t_{2g}^4 e_g^1$) with $S = 3/2$. Thus, it is likely that the spin state of $Sr_{1-x}Ca_xCoO_3$ remains to be the intermediate spin state for all compositions, even though the Co-O bond length slightly decreases upon the Ca substitution with $x$ up to 0.6. Considering the fact that the spin state of cubic perovskites $Sr_{1-x}Ba_xCoO_3$ has been also found to be independent of the Ba content[25], it is presumable that the intermediate spin state in the $Co^{4+}$-containing perovskites is stable over a wide range with respect to the Co-O bond length. Here we note that $SrFeO_3$ shows a similar field-induced transition from helimagnetic to conical spin state at 4 K and a positive Weiss temperature $\theta$. Figure 4(c) shows the $x$ dependence of the Weiss temperature $\theta$ for $Sr_{1-x}Ca_xCoO_3$, which is evaluated by the Curie-Weiss law. Upon the increment of the Ca content $x$, $\theta$ decreases monotonically while maintaining a positive value indicating the predominance of the FM interaction in all $x$ regions. In fact, $CaCoO_3$ of a cubic variant is reported to show FM behavior and the stability of FM state at high pressures is also predicted in the first-principles calculations[33]. We, hence, presume that $CaCoO_3$ adopts HM spin structure rather than G-type spin structure in the AFM phase as in the case of $SrFeO_3$ showing the pressure-induced-transition from a HM to FM state[34].

Figure 1(e) depicts the magnetic phase diagram of $Sr_{1-x}Ca_xCoO_3$ as a function of $x$ and $T$, where the transition temperature $T_c$ and $T_N$ are determined by the $M/H$-$T$ curves. The value of $T_c$ gradually decreases with increasing $x$, and the two types of FM phases are found in the cubic (FM1) and orthorhombic (FM2) structure phases. The drop of $T_c$ upon the change from FM1 to FM2 can be associated with the band-width narrowing caused by the first-order structural transition from the cubic to the orthorhombic phase. For $x = 0.8$, the AFM(HM) ground state emerges below the FM transition temperature. Finally, the FM transition was indiscernible for $x$=1.0. This phase diagram suggests that the FM and AFM interactions are competing with each other in $Sr_{1-x}Ca_xCoO_3$, and the FM interaction seems to be reduced by the Ca substitution. Here, the suppression of $T_c$ with increasing $x$ in the cubic phase is inconsistent with the naive expectation that the nearest neighbor FM interaction (double exchange interaction) should be enhanced via the band-width broadening accompanied by the decrease of the Fe-O bond length. In fact, the application of pressure for $SrCoO_3$ was found to enhance $T_C$.[14] For the origin of this $T_c$ suppression, two possible mechanisms can be considered: (i) Sr/Ca disorder similar to $Ln_{1/2}Ba_{1/2}MnO_3$ (Ln = La-Tb), where the A-site disorder suppresses $T_c$ in the vicinity of the multicritical composition, where FM and AFM phases are competing with each other[35,36] and (ii) local disorder of the orthorhombic structural distortion observed in the solid solution of perovskite-type oxides,[37,38] which should play important roles on the variation of the magnetic ground state as discussed below. .

Below, we discuss the possible origin of the AFM(HM) phase in terms of the change in the crystal structure and electronic states. As shown in Fig. 1, the unit cell volume decreases with increasing $x$, and the $CoO_6$ octahedra are largely tilted: the average Co-O-Co bond angle is about ~157°. First, the clear JT distortion is absent in the $CoO_6$ octahedra. In order to quantify the relative distortion of the deviation of the octahedra, we define the $\Delta_d$ parameter, denoting the deviation of Co-O distances $d_n$ with respect to the average <Co-O> value, as $\Delta_d = (1/6)\sum_{n=1,6}[(d_n - <d>)/<d>]^2$. For $CaCoO_3$, $\Delta_d$ is evaluated to be ~ $3 \times 10^{-5}$ which is substantially smaller than that of the JT system $LaMnO_3$ (~$5 \times 10^{-3}$),[39] but comparable to that of the non-JT system $RCoO_3$ (~$5 \times 10^{-5}$, R=Pr...Lu).[40]

The orthorhombic structural distortion influences the electronic properties as characterized by the $x$ dependence of $\rho$ and $C$. Figure 4(a) shows the temperature dependence of $\rho$, which is seemingly nonmetallic with a small magnitude less than 3 mΩcm at room temperature. The ratio of the electrical resistivity at 2 K to that at room temperature $\rho_{2K}/\rho_{300K}$ is enhanced especially for $x > 0.7$ as shown in Fig. 4(d), implying that the orthorhombic distortion changes the system to the incoherent metallic state, which is called a "bad metal", ubiquitously observed in the vicinity of the Mott insulator. The substantial orthorhombic distortion also affects the specific heat $C$. The $T^2$ dependence of $C/T$ is shown in Fig. 4(b). The Debye temperature $\theta_D$ and the electronic specific heat coefficient $\gamma$ were evaluated with the equation $C/T=\gamma + (12/5)\pi^4 NR\theta_D^{-3}T^2$ ($R$=8.31 J/mol K and $N$=5). $\theta_D$ changes discontinuously at the critical composition ($x = 0.6$), reflecting the structure transition from the cubic to orthorhombic phase, whereas $\gamma$ shows a gradual change with a kink at $x = 0.6$ (see Fig. 4(e)). The observation of large $\gamma$ of 40~50 mJmol$^{-1}$K$^{-2}$ in all compositions suggests that the effective mass of the conduction electron is enhanced by the electron correlation, as reported for the perovskite-type cobalt oxides $(Ca,Y)Cu_3Co_4O_{12}$.[41] As the Ca content increases in the orthorhombic phase, $\gamma$ gradually decreases presumably owing to the pseudo-gap formation causing the suppression of the density of states at the Fermi level, which is accompanied by the orthorhombic distortion. A recent band calculation with GGA+U for the orthorhombically distorted $CaCoO_3$ shows that only the majority $e_g$ bands are responsible for metallicity, while the minority $t_{2g}$ band opens a band gap[42]. These theoretical conjectures are compatible with the variation of the electronic properties of $Sr_{1-x}Ca_xCoO_3$, i.e., the incoherent metallic behavior with the low resistivity value (< 10 mΩcm) and suppression of $\gamma$ can be explained by the gap formation of the $t_{2g}$ bands while maintaining the itinerant $e_g$ bands. This is in contrast to the insulating ground state of $CaFeO_3$ exhibiting the charge disproportionation of $Fe^{4+}$ ions with $e_g$ bands near the Fermi level.

Finally, let us discuss the lattice dependent variation of the magnetic ground states of $Sr_{1-x}Ca_xCoO_3$. We can propose two possible mechanisms of the AFM(HM) ground state near $x = 1$. One is the double exchange mechanism for transition-metal oxides with the $e_g^1$ configuration and a negative $p$-$d$ charge-transfer energy $\Delta$.[6,43] Assuming that this model, in which the $pd$ hybridization plays an important role on the phase stability, holds for $Sr_{1-x}Ca_xCoO_3$ with the $e_g^1$ configuration and negative $\Delta$, the orthorhombic distortion shown in Fig. 1(d) presumably stabilizes the HM state rather than the FM state through the reduction of the $pd$ hybridization. The other is the frustration between the FM double exchange interaction $J_1$ and the AFM superexchange interaction $J_2$, the latter being comparable to the former when the orthorhombic distortion is significant as shown in Fig. 1(d).[43,45]. The orthorhombic distortion not only reduces the FM double exchange interaction due to the band narrowing but also enhances a next-nearest-neighbor (NNN) superexchange interaction through the decreasing in Co-O(1)-O(1)-Co exchange path, where the $t_{2g}$ electrons play an important role[46]. Given that the NNN superexchange interaction $J_2$ is AFM one, a noncollinear magnetic ground state is expected to manifest itself reflecting the competition with the nearest-neighbor FM interaction. To get further insight into the magnetic ground states, microscopic measurements using single crystals are indispensable.

The present work establishes the structure and magnetic phase diagram of $Sr_{1-x}Ca_xCoO_3$ and demonstrates a great potential of the $Co^{4+}$-containing perovskites as an interesting platform to explore a novel helimagnetic phase, which can be an important reference to study topological helimagnetic phases in $Fe^{4+}$-containing perovskites.

ACKNOWLEDGMENTS

The authors appreciate M. Takano, H. Sakai and J. Fujioka for the helpful suggestions. The authors thank M. Ashida, Y. Minowa, and M. Arai for experimental supports. This work is partly supported by JSPS, KAKENHI (Grants No. 19K14652, 21H01030 and 22H00343), JST PRESTO Hyper-nano-space design toward Innovative Functionality (JPMJPR1412), Asahi Glass Foundation, Murata Foundation, and Mazda Foundation. The powder XRD measurement was performed with the approval of the Photon Factory Program Advisory Committee (Proposal No. 2018S2-006).

REFERENCES

[1] J.E. Greedan, J. Mater. Chem. **11**, 37 (2001).

[2] U.K. Rössler, A.N. Bogdanov, and C. Pfleiderer, Nature **442**, 797 (2006).

[3] N. Kanazawa, S. Seki, and Y. Tokura, Adv. Mater. **29**, (2017).

[4] J.B. Torrance, P. Lacorre, A.I. Nazzal, E.J. Ansaldo, and C. Niedermayer, Phys. Rev. B **45**, 8209 (1992).

[5] N. Hayashi, T. Yamamoto, A. Kitada, A. Matsuo, K. Kindo, J. Hester, H. Kageyama, and M. Takano, J. Phys. Soc. Jpn. **82**, 113702 (2013).

[6] M. Mostovoy, Phys. Rev. Lett. **94**, 137205 (2005).

[7] H. Watanabe, J. Phys. Soc. Jpn. **12**, 515 (1957).

[8] S. Ishiwata, M. Tokunaga, Y. Kaneko, D. Okuyama, Y. Tokunaga, S. Wakimoto, K. Kakurai, T. Arima, Y. Taguchi, and Y. Tokura, Phys. Rev. B **84**, 054427 (2011).

[9] S. Ishiwata, T. Nakajima, J.-H. Kim, D.S. Inosov, N. Kanazawa, J.S. White, J.L. Gavilano, R. Georgii, K.M. Seemann, G. Brandl, P. Manuel, D.D. Khalyavin, S. Seki, Y. Tokunaga, M. Kinoshita, Y.W. Long, Y. Kaneko, Y. Taguchi, T. Arima, B. Keimer, and Y. Tokura, Phys. Rev. B **101**, 134406 (2020).

[10] M. Onose, H. Takahashi, H. Sagayama, Y. Yamasaki, and S. Ishiwata, Phys. Rev. Materials **4**, 114420 (2020).

[11] Y. Long, Y. Kaneko, S. Ishiwata, Y. Taguchi, and Y. Tokura, J. Phys. Condens. Matter **23**, 245601 (2011).

[12] R.H. Potze, G.A. Sawatzky, and M. Abbate, Phys. Rev. B **51**, 11501 (1995).

[13] M. Zhuang, W. Zhang, A. Hu, and N. Ming, Phys. Rev. B **57**, 13655 (1998).

[14] J.-Y. Yang, C. Terakura, M. Medarde, J.S. White, D. Sheptyakov, X.-Z. Yan, N.-N. Li, W.-G. Yang, H.-L. Xia, J.-H. Dai, Y.-Y. Yin, Y.-Y. Jiao, J.-G. Cheng, Y.-L. Bu, Q.-F. Zhang, X.-D. Li, C.-Q. Jin, Y. Taguchi, Y. Tokura, and Y.-W. Long, Phys. Rev. B **2**, 195147 (2015).

[15] J. Kuneš, V. Křápek, N. Parragh, G. Sangiovanni, A. Toschi, and A.V. Kozhevnikov, Phys. Rev. Lett. **109**, 117206 (2012).

[16] Y.W. Long, Y. Kaneko, S. Ishiwata, Y. Tokunaga, T. Matsuda, H. Wadati, Y. Tanaka, S. Shin, Y. Tokura, and Y. Taguchi, Phys. Rev. B **86**, 064436 (2012).

[17] A. Sukserm, U. Pinsook, T. Pakornchote, P. Tsuppayakorn-aek, W. Sukmas, and T. Bovornratanaraks, Comput. Mater. Sci. **210**, 111024 (2022).

[18] A.E. Bocquet, A. Fujimori, T. Mizokawa, T. Saitoh, H. Namatame, S. Suga, N. Kimizuka, Y. Takeda, and M. Takano, Phys. Rev. B **45**, 1561 (1992).

[19] M. Takano, N. Nakanishi, Y. Takeda, S. Naka, and T. Takada, Mater. Res. Bull. **12**, 923 (1977).

[20] P.M. Woodward, D.E. Cox, E. Moshopoulou, A.W. Sleight, and S. Morimoto, Phys. Rev. B **62**, 844 (2000).

[21] J. Matsuno, T. Mizokawa, A. Fujimori, Y. Takeda, S. Kawasaki, and M. Takano, Phys. Rev. B **66**, 193103 (2002).

[22] S. Kawasaki, M. Takano, and Y. Takeda, J. Solid State Chem. **121**, 174 (1996).

[23] T. Osaka, H. Takahashi, H. Sagayama, Y. Yamasaki, and S. Ishiwata, Phys. Rev. B **95**, 224440 (2017).

[24] H. Xia, J. Dai, Y. Xu, Y. Yin, X. Wang, Z. Liu, M. Liu, M.A. McGuire, X. Li, Z. Li, C. Jin, Y. Yang, J. Zhou, and Y. Long, Phys. Rev. Materials **1**, 024406 (2017).

[25] H. Sakai, S. Yokoyama, A. Kuwabara, J.S. White, E. Canévet, H.M. Rønnow, T. Koretsune, R. Arita, A. Miyake, M. Tokunaga, Y. Tokura, and S. Ishiwata, Phys. Rev. Materials **2**, 104412 (2018).

[26] F. Izumi and K. Momma, Solid State Phenom., **130**, 15 (2007).

[27] P.E. Blöchl, Phys. Rev. B **50**, 17953 (1994).

[28] P. Giannozzi, O. Andreussi, T. Brumme, O. Bunau, M. Buongiorno Nardelli, M. Calandra, R. Car, C. Cavazzoni, D. Ceresoli, M. Cococcioni, N. Colonna, I. Carnimeo, A. Dal Corso, S. de Gironcoli, P. Delugas, R.A. DiStasio Jr, A. Ferretti, A. Floris, G. Fratesi, G. Fugallo, R. Gebauer, U. Gerstmann, F. Giustino, T. Gorni, J. Jia, M. Kawamura, H.-Y. Ko, A. Kokalj, E. Küçükbenli, M. Lazzeri, M. Marsili, N. Marzari, F. Mauri, N.L. Nguyen, H.-V. Nguyen, A. Otero-de-la-Roza, L. Paulatto, S. Poncé, D. Rocca, R. Sabatini, B. Santra, M. Schlipf, A.P. Seitsonen, A. Smogunov, I. Timrov, T. Thonhauser, P. Umari, N. Vast, X. Wu, and S. Baroni, J. Phys. Condens. Matter **29**, 465901 (2017).

[29] N.E. Brese and M. O'Keeffe, Acta Crystallogr. B **47**, 192 (1991).

[30] P. Bezdicka, A. Wattiaux, J.C. Grenier, M. Pouchard, and P. Hagenmuller, Z. Anorg. Allg. Chem. **619**, 7 (1993).

[31] S. Balamurugan and E. Takayama-Muromachi, J. Solid State Chem. **179**, 2231 (2006).

[32] S. Balamurugan, K. Yamaura, A.B. Karki, D.P. Young, M. Arai, and E. Takayama-Muromachi, Phys. Rev. B **74**, 172406 (2006).

[33] A. Sukserm, U. Pinsook, T. Pakornchote, P. Tsuppayakorn-aek, W. Sukmas, and T. Bovornratanaraks, Comput. Mater. Sci. **210**, 111024 (2022).

[34] T. Kawakami, and S. Nasu J. Phys. Condens. Matter **17**, S789 (2005).

[35] J.P. Attfield, Chem. Mater. **10**, 3239 (1998).

[36] D. Akahoshi, M. Uchida, Y. Tomioka, T. Arima, Y. Matsui, and Y. Tokura, Phys. Rev. Lett. **90**, 177203 (2003).

[37] A. Vegas, M. Vallet-Regí, J.M. González-Calbet, and M.A. Alario-Franco, Acta Crystallogr. B **42**, 167 (1986).

[38] J. Rodríguez-Carvajal, M. Hennion, F. Moussa, A.H. Moudden, L. Pinsard, and A. Revcolevschi, Phys. Rev. B **57**, R3189 (1998).

[39] J.A. Alonso, M.J. Martínez-Lope, M.T. Casais, and M.T. Fernández-Dáz, Inorg. Chem. **39**, 917 (2000).

[40] J.A. Alonso, M.J. Martínez-Lope, C. de la Calle, and V. Pomjakushin, J. Mater. Chem. **16**, 1555 (2006).

[41] I. Yamada, S. Ishiwata, I. Terasaki, M. Azuma, Y. Shimakawa, and M. Takano, Chem. Mater. **22**, 5328 (2010).

[42] S. Nazir, J. Alloys Compd. **732**, 187 (2018).

[43] Z. Li, R. Laskowski, T. Iitaka, and T. Tohyama, Phys. Rev. B **85**, 134419 (2012).

[44] P.-G. de Gennes, Phys. Rev. **118**, 141 (1960).

[45] J.-H. Kim, A. Jain, M. Reehuis, G. Khaliullin, D.C. Peets, C. Ulrich, J.T. Park, E. Faulhaber, A. Hoser, H.C.

Walker, D.T. Adroja, A.C. Walters, D.S. Inosov, A. Maljuk, and B. Keimer, Phys. Rev. Lett. **113**, 147206 (2014).

[46] T. Kimura, S. Ishihara, H. Shintani, T. Arima, K.T. Takahashi, K. Ishizaka, and Y. Tokura, Phys. Rev. B **68**, 060403 (2003).

Table I.  Refined structural parameters and the bond-valence sum (BVS) of $Co^{4+}$ for $Sr_{1-x}Ca_xCoO_3$. BVS is calculated by $\sum_i exp[(r_0 - r_i)/0.37]$, where $r_0$=1.75 .

| $x$ | 0.2 | 0.4 | 0.6 | | 0.8 | 1.0 |
|---|---|---|---|---|---|---|
| space group | *Pm-3m* | *Pm-3m* | *Pm-3m* | *Pbnm* | *Pbnm* | *Pbnm* |
| a (Å) | 3.8255(1) | 3.8040(1) | 3.7857(4) | 5.3264(8) | 5.3152(8) | 5.2664(5) |
| b (Å) | | | | 5.3109(9) | 5.3139(5) | 5.2929(6) |
| c (Å) | | | | 7.548(1) | 7.500(1) | 7.4354(6) |
| V/Z (Å$^3$) | 55.984 | 55.047 | 54.253 | 53.383 | 52.960 | 51.815 |
| Co-O(1) (Å) | 1.91276(6) | 1.90201(7) | 1.8928(2) | 1.78(3) ×2  2.07(2) ×2 | 1.938(16) ×2  1.870(16) ×2 | 1.914(8) ×2  1.905(8) ×2 |
| Co-O(2) (Å) | | | | 1.888(1) ×2 | 1.919(3) ×2 | 1.890(2) ×2 |
| <Co-O-Co>(°) | 180 | 180 | 180 | 155.8(9) ×4  178(4) ×2 | 161.4(9) ×4  155.4(8) ×2 | 155.7(5) ×4  159.1(7) ×2 |
| Co Valence (BVS) | 3.86 | 3.98 | 4.08 | 4.08 | 3.92 | 3.97 |
| $R_{wp}$ (%) | 0.818 | 1.031 | 1.114 | | 1.384 | 1.961 |
| S | 1.385 | 2.563 | 1.293 | | 2.519 | 3.531 |

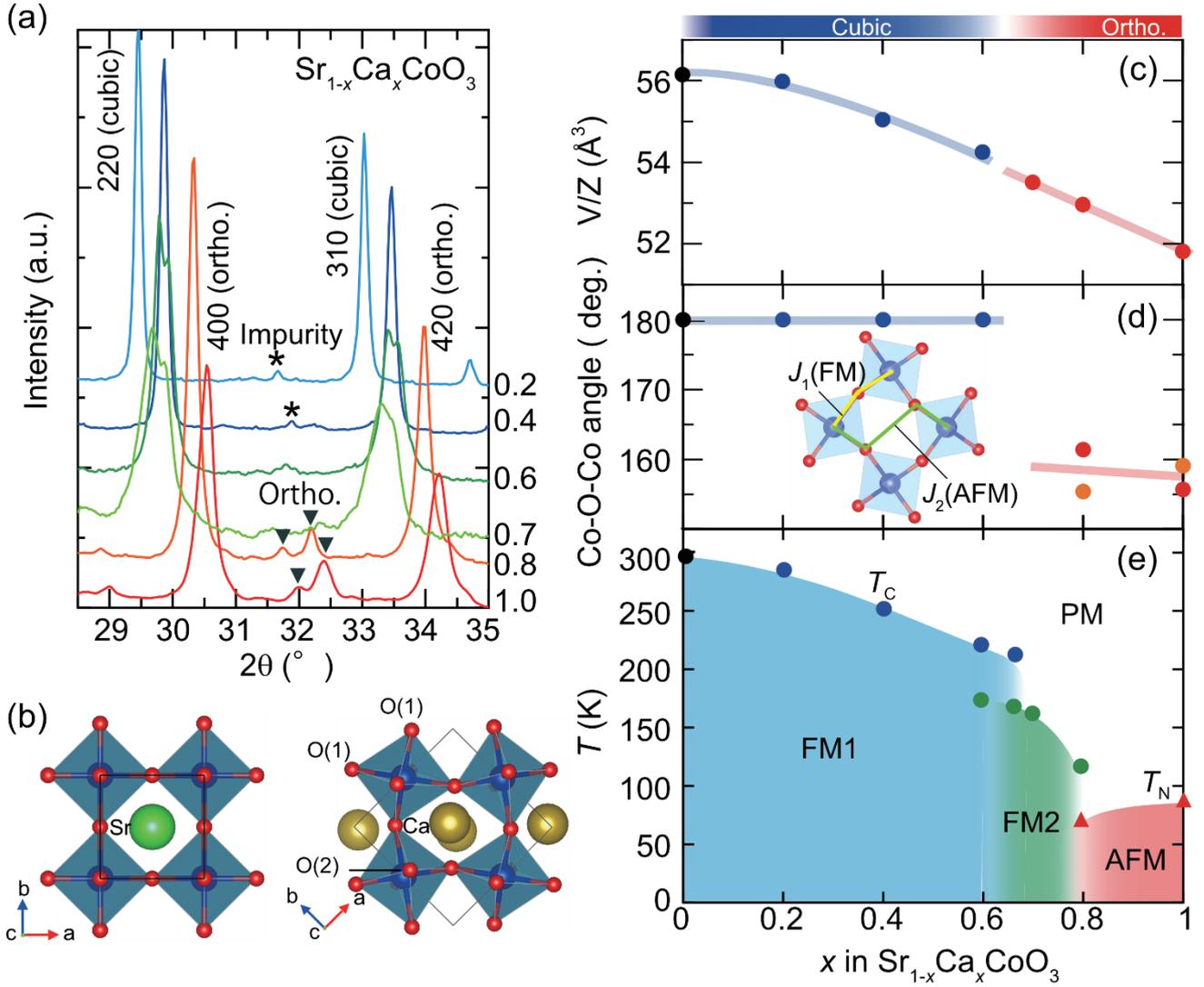

Figure 1. (a) Synchrotron powder XRD patterns for $Sr_{1-x}Ca_xCoO_3$. (b) Crystal structures of the cubic-perovskite $SrCoO_3$ and orthorhombic-perovskite $CaCoO_3$ projected along the $c$ axis. The blue, green, and yellow spheres correspond to Co, Sr, and Ca ions, respectively. (c) Unit cell volume $V$ divided by number of atoms per unit cell $Z$, and (d) average Co-O-Co bond angle as a function of $x$ for $Sr_{1-x}Ca_xCoO_3$. Black dots in (c) and (d) are the data of the single crystalline sample[11]. The inset in Fig. 1(d) shows the schematic illustration of the nearest-neighbor (NN) interaction $J_1$ and next-nearest-neighbor (NNN) interaction $J_2$ for the orthorhombic structures. (e) Magnetic phase diagram of $Sr_{1-x}Ca_xCoO_3$ with abbreviations: FM for ferromagnetic, AFM for antiferromagnetic. $T_c$ of the single crystal at $x=0$ is taken from Ref. [11].

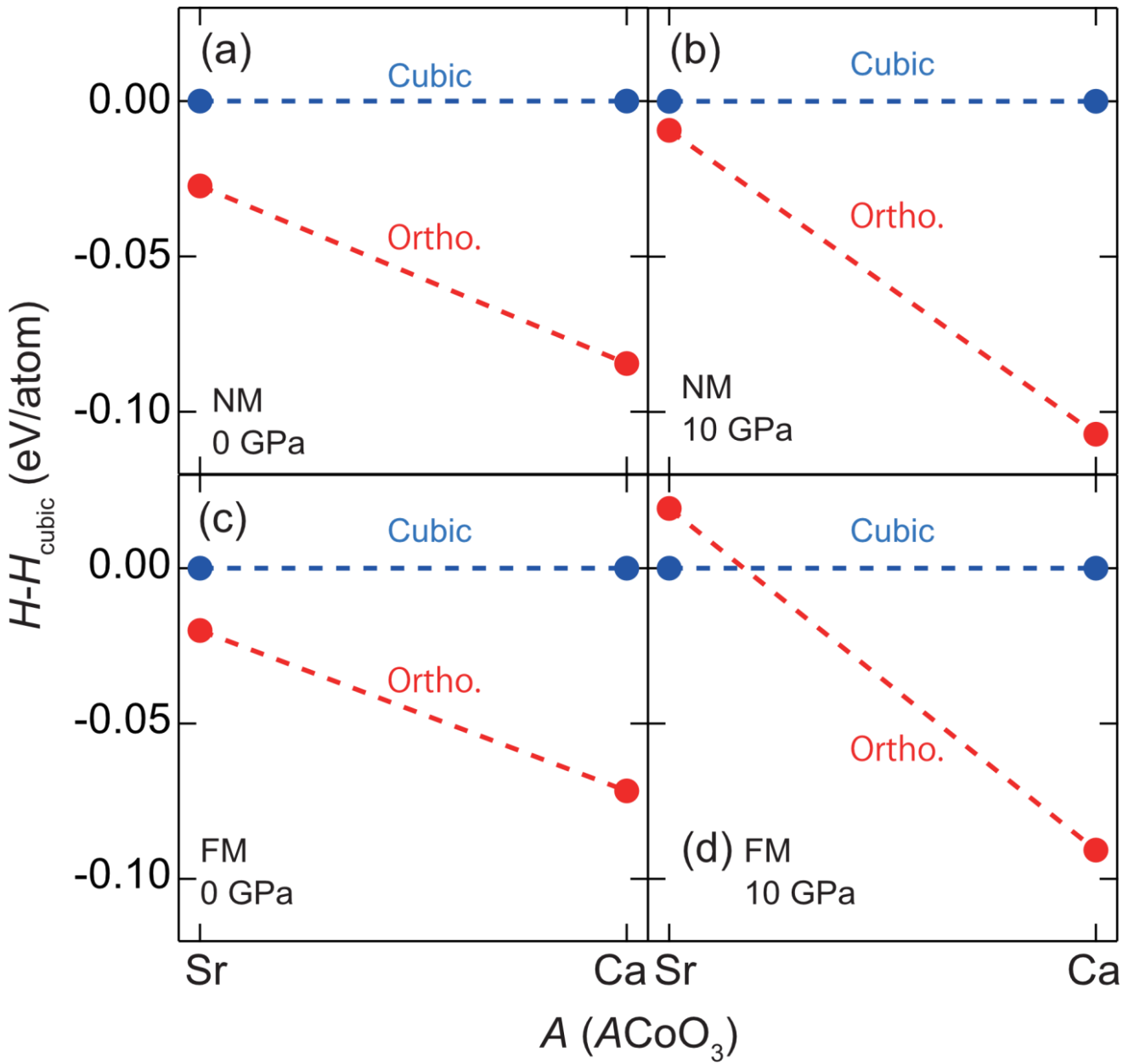

Figure 2. (a) Enthalpy difference between the orthorhombic (*Pbnm*) and cubic (*Pm-3m*) phases of $SrCoO_3$ and $CaCoO_3$ calculated for (a,b) nonmagnetic (NM) and (c,d) ferromagnetic (FM) phases at (a,c) 0 GPa and (b,d) 10 GPa.

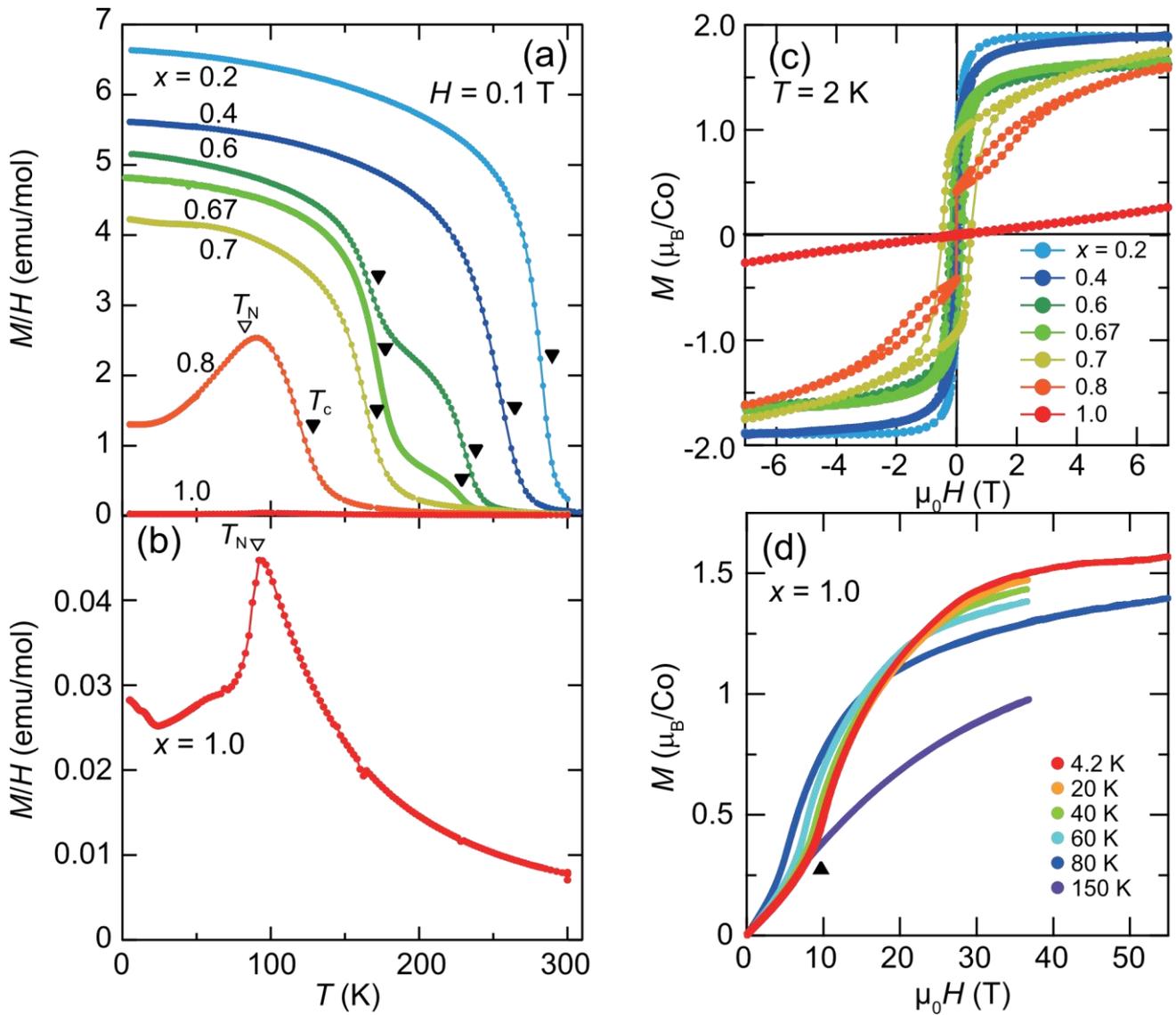

Figure 3. (a) Temperature dependence of the magnetization divided by the applied magnetic field $M/H$ for $Sr_{1-x}Ca_xCoO_3$. (b) $M/H$ for $x=1.0$ as a function of temperature. (c) Magnetic field dependence of the magnetization for $Sr_{1-x}Ca_xCoO_3$ at 2 K. (d) Magnetic field dependence of the magnetization for $x=1.0$ up to 55 T at selected temperatures.

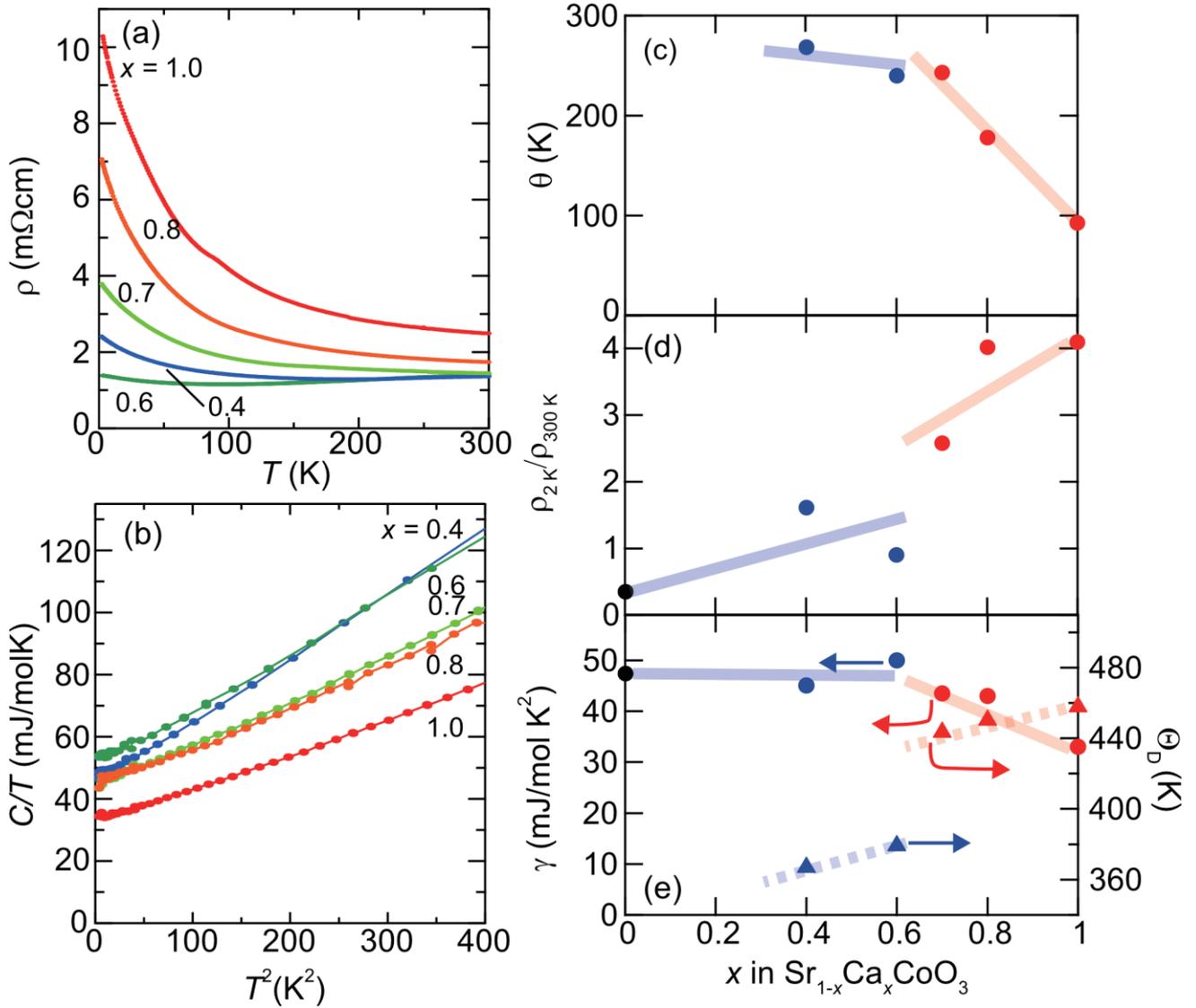

Figure 4. (a) Temperature dependence of $\rho$ and (b) $T^2$ dependence of $C/T$ for $Sr_{1-x}Ca_xCoO_3$. (c) Weiss temperature $\theta$ and (d) the ratio of the electrical resistivity at 2 K to that at room temperature $\rho_{2K}/\rho_{300K}$ as a function of $x$. (e) Electronic specific heat coefficient $\gamma$ (left axis) and Debye temperature $\theta_D$ (right axis) as a function of $x$. Black filled circles at $x = 0$ in (d) and (e) are the reported data of the polycrystalline sample.[32]